\documentclass[doublecol]{epl2}

\usepackage{epsfig}
\usepackage{amsmath}
\usepackage{bm}
\usepackage{float}
\usepackage{graphicx}
\usepackage[linkcolor=black,urlcolor=blue,citecolor=blue,colorlinks=true]{hyperref}
\usepackage{xcolor}

\usepackage[normalem]{ulem}

\title{Observation of pseudogap in MgB$_2$}

\shorttitle{Pseudogap in MgB$_2$}

\author{Patil S.,$^{1,2}$ Medicherla V. R. R.,$^1$ Ali Khadiza,$^1$ Singh R. S.,$^1$ Manfrinetti P.,$^{3,4}$ Wrubl F.,$^3$ Dhar S. K.$^1$ \and Maiti Kalobaran$^1${\footnote{Corresponding author: kbmaiti@tifr.res.in}}}

 \shortauthor{Patil {\it et al.}}

\institute{$^1$Department of Condense Matter Physics and Materials' Science, Tata Institute of Fundamental Research, Homi Bhabha Road, Colaba, Mumbai-400005, India\\
$^2$Department of Physics, Indian Institute of Technology (Banaras Hindu University), Varanasi-221005, India\\
$^3$Department of Chemistry, Universit\'{a} di Genova, 16146 Genova, Italy\\
$^4$Institute SPIN-CNR, 16152 Genova, Italy.}


\abstract{Pseudogap phase in superconductors continues to be an outstanding puzzle that differentiates unconventional superconductors from the conventional ones (BCS-superconductors). Employing high resolution photoemission spectroscopy on a highly
dense conventional superconductor, MgB$_2$, we discover an interesting scenario. While the spectral evolution close to the Fermi energy is commensurate to BCS descriptions as expected, the spectra in the wider energy range reveal emergence of a pseudogap much above the superconducting transition temperature indicating apparent
departure from the BCS scenario. The energy scale of the pseudogap is comparable to the energy of $E_{2g}$ phonon mode responsible for superconductivity in MgB$_2$ and the pseudogap can be attributed to the effect of electron-phonon coupling on the electronic structure. These results reveal a scenario of the emergence of the superconducting gap within an electron-phonon coupling induced pseudogap.}

\begin{document}

\maketitle

\section{Introduction}

One of the most intriguing aspects of high temperature superconductivity is the observation of \textcolor[rgb]{0.00,0.00,1.00}{a} pseudogap above the superconducting transition temperature, $T_c$
\cite{Vishik2010,Timusk1999,Hufner2008}. In the pseudogap phase, the spectral density of states (SDOS) exhibits an energy gap at the Fermi level at some $k$-points around the antinode in the Brillouin zone above $T_c$, which persists upto another temperature scale, $T^{*}$ \cite{Norman1998}. The Fermi surface gradually grows with the increase in temperature around the nodal point as an arc called `Fermi arc' and the full Fermi surface appears at $T^*$. The existence of this phase is considered to be one of the major features distinguishing unconventional superconductors from the conventional ones.

Two schools of thought exist on the origin of the pseudogap phase in unconventional superconductors. One scenario proposes the pseudogap to represent paired electronic states without superconducting coherence \cite{Randeria1998,Franz2007,Emery1995,Curty2003}. In this picture, the superconducting transition temperature, $T_c$, is the temperature where phase coherence amongst the wave functions of the electron pairs sets in, giving rise to resistance less electrical behaviour. The incoherent paired electronic states above $T_c$ are the precursor to superconductivity. In the second scenario, the pseudogap is disconnected from superconductivity and is believed to arise due to other competing interactions such as a hidden order, spin fluctuations, {\it etc.} \cite{Moon2010,Moon2009,Sachdev2010}. These competing interactions are believed to culminate into a Quantum Critical Point (QCP) deep within the superconducting dome in the phase diagram.

The conventional superconductors are expected not to possess such complexity. Some recent studies, however, show signature of pseudogap even in conventional superconductors
\cite{Medicherla2007,Chainani2001,Yokoya2002,Sacepe2010,Mondal2011,Thakur2013}.
In most of these studies, however, the appearance of pseudogap was often attributed to crystallographic disorder and/or reduced dimensionality as commonly observed in non-superconducting disordered metals \cite{Altshuler1979,Kobayashi2007,Sarma1998,Lee1985,MaitiPRB07,MaitiEPL07}. Thus, the
connection of the pseudogap (presumably arising due to extraneous effects) and its influence, if any, to superconductivity in the conventional superconductors remains obscured. It is, thus, desirable to search for a superconductor in which the spectral
function at the Fermi energy reflects an intrinsic electronic structure free from dimensionality factors and/or disordered lattice potential. In this work, we choose a 3-dimensional bulk compound, MgB$_2$, which is a $high$ temperature conventional superconductor and not $nominally$ disordered. We present our high resolution photoemission spectroscopic results on specially prepared highly compact MgB$_{2}$ \cite{Budko2001,Nagamatsu2001} as a function of temperature, which reveals signature of pseudogap in MgB$_2$. The relatively high $T_c$ ($~$39 K; $\Delta T_c$ =
0.2 K) of MgB$_2$ presents a large temperature range to study the pseudogap physics in detail using traditional photoemission spectrometers equipped with liquid helium cryomanipulator, which is a distinct advantage over other conventional systems.

\section{Experimental details}

High resolution photoemission spectroscopic (HRPES) measurements were performed on highly compacted samples to minimize the extraneous effects due to impurities, grain-boundary oxides (particularly MgO) and extra phases, disorder, defect {\em etc.} The novel synthesis method patented by our group (apt to prepare bulks) is a concomitant grain growth method via direct synthesis from pure B (99.96$\%$) and Mg (99.99$\%$) put into outgassed Ta crucibles, sealed by arc welding in pure argon and subsequently, an annealing at $950^{o}C$ for 3 days was followed \cite{Manfrinetti:misc}. The final products are brown-blackish and was extremely compact with grains highly connected by polytwinning, which could not be used to prepare samples in single crystalline form. The samples possess density value of the order of 2.3 g/cm$^3$ with respect to a theoretical value of 2.6 g/cm$^3$ calculated on the basis of the unit-cell volume. Magnetic and transport measurements exhibit that the pristine properties are optimal, $T_c$ = 39.2 K and the superconducting transition is {\it very sharp} ($\Delta T_c$ = 0.2 K). The quality of the MgB$_{2}$ phase was studied by $x$-ray powder diffraction and scanning electron microscope - energy dispersive analysis of $x$-rays (SEM-EDX)
measurements. No free Mg or spurious phases were detected.

For photoemission spectroscopic (PES) measurements, the samples were fractured/scraped in-situ (pressure $< 2\times10^{-11}$~Torr) at each temperature and cleanliness was confirmed by the absence of impurity feature in both  $x$-ray photoemission spectroscopy (XPS) and ultraviolet photoemission spectroscopy (UPS) measurements. Reproducibility of the spectra was ascertained after each cycle of surface preparation. Photoemission measurements were carried out using a monochromatic Al $K\alpha$ $x$-ray source ($h\nu$ = 1486.6 eV) for $x$-ray photoemission spectroscopy (XPS) and He {\scriptsize
I}$\alpha$ photon source ($h\nu$ = 21.2 eV) for ultra violet photoemission spectroscopy (UPS) employing a spectrometer equipped with SES2002 Gammadata Scienta analyzer with an energy resolution set to 300~meV for XPS and 1.4 meV for UPS measurements. XPS was
used to monitor the growth of impurities on the sample surface. The temperature variation was achieved using an open cycle He cryostat from Advanced Research Systems, USA. The lowest temperature of 5 K could be achieved by pumping helium at the exhaust.

\section{Results}

\begin{figure}[htbp]
\centering
\includegraphics[keepaspectratio,width=\linewidth,height=0.45\textheight]{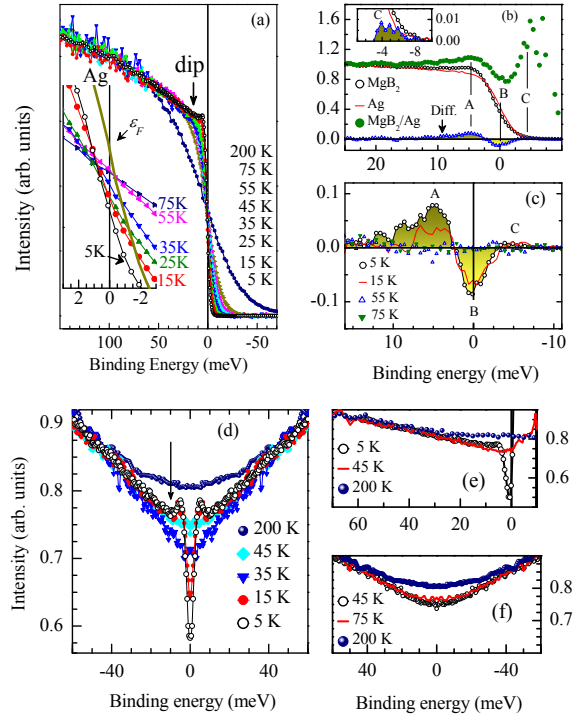}
\caption{(a) Valence band spectra of MgB$_2$ at various
temperatures. The inset shows near \textbf{the} $\epsilon_F$ part in an expanded
energy scale. (b) Valence band spectra of MgB$_2$ (open circles) and
Ag (line) at 5 K, and their ratio (solid circles). Difference of
MgB$_2$ and Ag spectra at 5 K are shown by triangles. The inset
shows enlarged above $\epsilon_F$ part of the different spectra. (c)
Difference of MgB$_2$ and Ag valence band spectra at various
temperatures. (d) Symmetrized spectral function of MgB$_2$ valence
band spectra at different temperatures. (e) Spectral DOS obtained by
the division of resolution broadened Fermi-Dirac function. (f)
Symmetrized spectral DOS above T$_C$.}
\end{figure}

In Fig. 1(a), we show the high resolution He {\scriptsize I}$\alpha$ spectra collected at different temperatures across $T_c$. Signature of a distinct coherent peak close to the Fermi level, $\epsilon_F$, followed by a dip at around 10 meV shown by an arrow is observed in the 5 K spectrum.  Below $T_c$, the spectral weight at $\epsilon_F$
gradually shifts towards higher binding energies, resulting in the opening up of an energy gap. The lowest temperature case is shown in Fig. 1(b), where we show the high resolution 5~K He {\scriptsize I}$\alpha$ spectrum close to $\epsilon_F$ superimposed over the Ag spectrum collected under identical experimental conditions. The difference between the two spectra is clearly manifested in the figure. The leading edge of the MgB$_2$ spectrum appears at higher binding energy relative to the Ag spectrum implying the formation of the superconducting gap. In order to elucidate the spectral weight changes more clearly, we show the difference between the spectral intensity of MgB$_2$ and Ag by open triangles. There are {\it three} features in the difference spectrum as denoted by A, B and C - the existence of the feature C is more apprehensible in an expanded intensity scale shown in the inset. A division of the MgB$_2$ spectrum by the Ag spectrum that takes care of the Fermi-Dirac distribution induced spectral suppression across $\epsilon_F$ reveals more distinctly the signatures of A, B and C - both A \& C
are peaks, and B is a valley/dip representing the superconducting gap - a {\it peak-dip-peak} structure typical for a superconductor in the superconducting phase. The temperature evolution of the difference spectra shown in Fig. 1(c), demonstrates the gradual filling up of the dip and smearing of the peaks A \& C with increasing temperature. Above $T_c$, all these features are smeared out. These spectral changes demonstrate the vanishing of the superconducting gap above $T_c$ commensurate to the predictions of the Bardeen-Cooper-Schrieffer (BCS) theory \cite{Chainani2001,Bardeen1957,Bardeen1957a,Reinert2000}.

The Fermi function broadening superimposed on the spectral functions renormalizes the changes expected from the many-body physics of the system. Thus, we extracted the spectral density of states (SDOS) by symmetrizing the experimental spectra, which often provides a good representation of SDOS, and has been found to be very helpful in
analyzing the superconducting gap/pseudogap in cuprate superconductors as well as in other systems \cite{Norman1998,MaitiEPL07,bairo3}. The effect of resolution broadening would be minimal due to high energy resolution of 1.4 meV employed in the experimental technique.

\begin{figure}[]
\centering
\includegraphics[keepaspectratio,width=\linewidth,height=0.45\textheight]{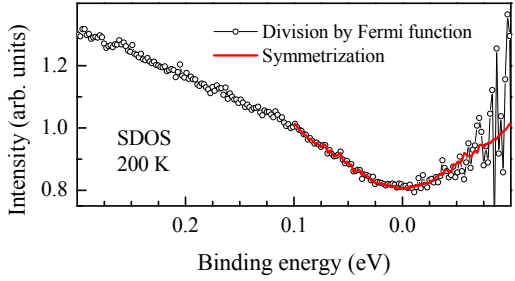}
\vspace{-40ex}
 \caption{Comparison of the spectral density of states at 200 K obtained by symmetrization (line) and via division by the resolution broadened Fermi-Dirac function (symbols).}
\end{figure}

Since the symmetrization method implicitly assume an electron-hole symmetry in the spectral function, such method of extraction fails to capture particle-hole asymmetry, if there is any in the systems. Thus, the procedure of extraction of SDOS has been verified in Fig. 2 by comparing the SDOS obtained by symmetrization (line) with the spectral function obtained by the division of the resolution broadened Fermi-Dirac distribution function (open circles). The later procedure does not suffer from such effect and helps to extract the SDOS from high resolution data quite reliably. The figure shows an identical representation of SDOS obtained from both the procedures providing confidence on the analysis process adopted here. Thus obtained SDOS are shown in Fig. 1(d) exhibiting additional distinct signature of {\it peak-dip-peak} structure at low temperatures that gradually smears out with the increase in temperature. Here, the dip refers to the valley observed at 10 meV between the coherent feature and the spectral intensities at higher binding energies as shown by an arrow in Fig. 1(d) and 1(a).

\begin{figure}[]
\centering
\includegraphics[keepaspectratio,width=\linewidth,height=0.45\textheight]{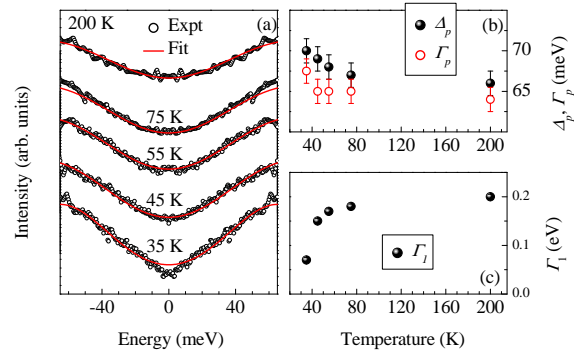}
\vspace{-32ex}
 \caption{(a) Fit of the SDOS at different temperatures. (b) Pseudogap and
the lifetime of the phonon coupled carriers. (c) Single particle
scattering rate with temperature.}
\end{figure}

Interestingly, the curves representing spectral intensities at different temperatures start to deviate from each other at around 50~meV binding energy [see Figs. 1(d) \& 1(f)]. Such spectral evolution with decreasing temperature is reminiscent of a pseudogap
formation in metals. We observe that SDOS at $\epsilon_F$ decreases monotonically with decreasing temperature in the normal phase ($T > T_c$) as shown separately in Fig. 1(f) for clarity. The {\it peak-dip-peak} structure below $T_c$ corresponding to the
superconducting gap appears to emerge from the pseudogap at $T > T_c$. In order to explore the evolution of this pseudogap with temperature, we simulated the spectral function assuming a phenomenological self energy of the form, $\Sigma_k(\epsilon) =
-i\Gamma_1 + \Delta_p^2/[(\epsilon + i\Gamma_p) +\epsilon_k]$, where the second term represents the effect due to electron-phonon coupling. $\Gamma_1$ is the single particle scattering rate. $\Delta_p$ and $\Gamma_p$ represent the pseudogap and lifetime of
the phonon-dressed electrons in the system. The electron-phonon coupling constant, $\lambda$ can be expressed as, $\lambda = D_{\epsilon_F}<I^2>/M<\omega^2>$, where $D_{\epsilon_F}$ is the density of states (DOS) at $\epsilon_F$ per spin per atom, $<I^2>$ is the square of the averaged electron-ion matrix element involving lattice vibrations, $M$ is the atomic mass, and $<\omega^2>$ is the average of the square of the phonon frequencies \cite{mcmillan}. The product $M<\omega^2>$ does not depend on the mass, but on the force constants, while $\eta = D_{\epsilon_F}<I^2>$, known as the Hopfield factor, is purely an electronic property and depends on spectral function at $\epsilon_F$. $\Delta_p$ provides a measure of the pseudogap due to the renormalized spectral function at $\epsilon_F$ and $both$ $\Gamma_p$ \& $\Delta_p$ depend on $\lambda$.

The simulated SDOS shown by solid line in Fig. 3(a) provides an excellent representation of the experimental SDOS. A small deviation at $\epsilon_F$ appears at 35 K, which is below $T_c$ due to the emergence of the superconducting gap. The corresponding parameters
are shown in Fig. 3(b) and 3(c). The single particle scattering rate is found in the range of 0.1 - 0.2 eV as expected. The value of $\Gamma_p$ is about 65 - 70 meV indicating finite lifetime of the electron-phonon coupled many body states. The pseudogap of about 65-70 meV found here matches remarkably well with the energy of the $E_{2g}$ phonon mode found in Raman studies, which is believed to be responsible for the BCS-type superconductivity in this system \cite{Osborn2001,Quilty2002}. Small increase in $\Delta_p$ and $\Gamma_p$ with the decrease in temperature may be attributed to the
increase in spring constant at lower temperatures.

\begin{figure}[htbp]
\centering
\includegraphics[keepaspectratio,width=\linewidth,height=0.45\textheight]{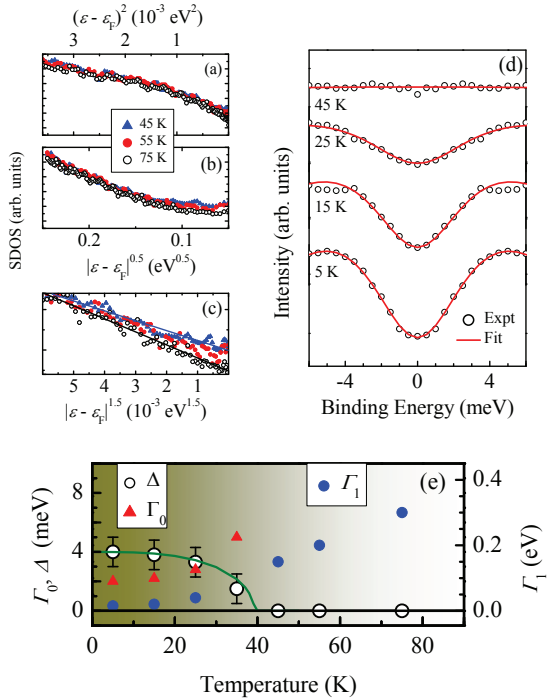}
\caption{SDOS plotted as a function of (a)
$(\epsilon-\epsilon_F)^2$, (b) $(\epsilon-\epsilon_F)^{0.5}$ and (d)
$(\epsilon-\epsilon_F)^{1.5}$ respectively. (d) The fit of the near
$\epsilon_F$ part of the spectra. (e) Fitting parameters as a
function of temperature.}
\end{figure}

In order to verify the influence of disorder in the spectral function, we plot SDOS as a function $(\epsilon-\epsilon_F)^\alpha$ with different numerical values of $\alpha$ in Fig. 4(a). In the case of disordered metals, it is well known that the SDOS varies as
$(\epsilon-\epsilon_F)^{0.5}$ \cite{Altshuler1979,Kobayashi2007,Sarma1998}. We chose to study the SDOS at temperatures above $T_c$ in order to exclude the influence of BCS spectral function over the SDOS. In Fig. 4(a), $\alpha$ = 2 gives rise to a non-linear behaviour with a downward curvature and $\alpha$ = 0.5 results in an upward curvature (see Fig. 4(b)). Evidently, the optimal $\alpha$ value lies between 0.5 and 2 indicating a deviation from the simple Fermi liquid behaviour as found in other systems \cite{Thakur2013,MaitiEPL07,y2ir2o7,epl-Cr}. Interestingly, for $\alpha$ = 1.5, a linear behaviour of SDOS is observed (see Fig. 4(c)), which has often been attributed to the influence of the coupling of electron to collective excitations such as magnon in magnetic materials \cite{y2ir2o7,Irkhin1990,Irkhin1993}. The observation of 1.5
exponent in non-magnetic MgB$_2$ indicates a link of the spectral functions and collective excitations.

The spectral function close to $\epsilon_F$ is simulated using a phenomenological self energy as used earlier \cite{Norman1998,Thakur2013} and superimposed over the symmetrized
spectra in Fig. 4(b). Since the spectral change due to the superconductivity occurs at an energy scale of about 5~meV (quasiparticle binding energy) in Mg$B_2$, we use an energy range of about $\pm$10~meV across $\epsilon_F$ of the experimental spectra for fitting. Fig. 4(d) shows the results of the simulations with remarkably good representation of the experimental spectra for a realistic sets of parameters shown in Fig. 4(e) \cite{Norman1998,Takahashi2001}. The temperature evolution is captured by the variation of the parameters reasonably well. The single particle scattering rate, $\Gamma_1$ = 0.2 eV found in this energy range is similar to the value found at higher energy range at high temperatures. The superconducting gap, $\Delta$ gradually decreases with the increase in temperature and vanishes at and above $T_c$. The temperature dependence \cite{choi_nature} can be represented by $\Delta(T) = \Delta(0)\sqrt{1-(T/T_c)^\beta}$ with $\beta$ = 3.0 and $T_c$ = 39.2 K. This indicates a typical BCS-type superconductivity in this material with $\Delta(0)$ = 4 meV and $T_c$ = 39.2 K consistent with the transport data
\cite{Takahashi2001,choi_nature,Tsuda2001,Tsuda2003,Tsuda2002}.

\section{Discussion}

From the above experimental results, it is clear that the superconducting gap in MgB$_2$ vanishes at the superconducting transition temperature as expected in a typical BCS-type
superconductor. However a distinct pseudogap survives to temperatures much higher than its $T_c$, which is unusual for a BCS superconductor, calling for a new scenario in these conventional systems. Revelation of such scenario was possible due to the high quality of the compacted MgB$_2$ sample \cite{Manfrinetti:misc} and high resolution of the technique employed. It is difficult to associate the pseudogap in this compound to the survival of Cooper pairs above $T_c$ as observed in cuprate superconductors since the energy scales are very different in MgB$_2$ while in cuprates, the pseudogap above $T_c$ scales well with the superconducting gap leading to the claim that Cooper pairs exist above $T_c$. Absence of $(\epsilon-\epsilon_F)^{0.5}$ dependence of the SDOS implies that the role of crystalline disorder can also be ruled out as the cause of the pseudogap.

The binding energy at which SDOS starts to decrease with temperature and the size of the pseudogap corroborate well with the $E_{2g}$ phonon excitations involving vibration of B-atoms in the B-sublattice \cite{Osborn2001,Quilty2002}, which is believed to cause superconductivity in this material. Thus, the pseudogap can be attributed to the electrons coupling to collective excitations of the crystal lattice (phonons) as also found in various other systems \cite{Medicherla2007,Chainani2001,Yokoya2002,Thakur2013}. Some experiments on cuprates also exhibit a high energy pseudogap ($\sim$~0.1 eV in addition to ones associated with the Fermi arcs) scaling with the characteristic temperature/energy scales arising due to the antiferromagnetic correlations or short range antiferromagnetic order in the system \cite{Ino1998,Sato1999,Sato2000}. It is to be noted that the presence of electron phonon interaction is theoretically explained in the normal state of MgB$_2$ \cite{Kong}. Furthermore, the continuous temperature evolution (without any anomaly) of the frequency across $T_c$ and linewidth of the Raman spectra from the $E_{2g}$ phonon mode, possibly lends further credence to the existence of electron phonon coupling above $T_c$ in MgB$_2$ \cite{Martinho}.

An important contrast with the cuprate superconductors is that there are two kinds of pseudogap present in cuprates - (i) the high energy one ($\sim$~0.1 eV) is associated to the antiferromagnetic order and (ii) the ones forming Fermi arcs appear to merge with the
superconducting gap i.e. energy scale of the second case is similar to the energy scale of the superconducting gap. In MgB$_2$, the energy scale over which the pseudogap is observed is much larger ($\geq$50~meV) than the energy scale over which superconducting gap is observed ($\sim$4-6~meV) and is similar to the first case in cuprates. Our observation makes an impression that the superconducting gap emerges within the pseudogap formed due to electron phonon coupled objects. The pseudogap consisting of
electrons coupled to phonons with finite lifetime presumably provides the necessary platform for electrons within a certain energy range to form Cooper pairs and cause superconductivity.

\section{Conclusions}

In summary, we have studied the electronic structure of MgB$_2$ employing high resolution photoemission spectroscopy. Our high resolution valence band photoemission results on a highly dense sample show the
existence of a pseudogap in MgB$_2$ below 200~K and a superconducting gap below $T_c$. The energy scale over which the pseudogap forms ($\sim$65-70~meV) is significantly larger than the energy scale of the superconducting gap ($\sim$4-6~meV) and corroborates well with the $E_{2g}$ phonon excitations of the system. A probable picture of superconductivity in MgB$_2$ can be conjectured to be the one where the electron pairs causing superconductivity emerges from the electron-phonon coupled species already formed above $T_c$.

\section{Acknowledgements}

S.P. acknowledges financial support from the Council of Scientific and Industrial Research, Govt. of India. K. M. acknowledges financial assistance from the Department of Science and Technology, Govt. of India (Swarnajayanti Fellowship and J.C. Bose Fellowship) and Department of Atomic Energy, Govt. of India.

\end{document}